\documentclass[runningheads]{llncs}
\usepackage{graphicx}
\usepackage{todonotes}
\usepackage{subfiles}
\usepackage{caption}
\usepackage{subcaption}
\usepackage{hyperref}
\usepackage{booktabs}
\hypersetup{
    colorlinks=true,
    linkcolor=blue,
    filecolor=magenta,      
    urlcolor=cyan,
}
\usepackage{amsmath}
\usepackage{adjustbox}
\begin{document}
%
\title{Detecting Hypo-plastic Left Heart Syndrome in Fetal Ultrasound via Disease-specific Atlas Maps}
\titlerunning{Prenatal HLHS prediction with Atlas-ISTNs}
\author{Samuel Budd\inst{1}\orcidID{0000-0002-9062-0013},
Matthew Sinclair\inst{1},
Thomas Day\inst{2,3},
Athanasios Vlontzos\inst{1},
Jeremy Tan\inst{1},
Tianrui Liu\inst{1},
Jacqueline Matthew\inst{2,3},
Emily Skelton\inst{2,3,4},
John Simpson\inst{2,3},
Reza Razavi\inst{2,3},
Ben Glocker\inst{1},
Daniel Rueckert\inst{1,5},
Emma C. Robinson\inst{2},
Bernhard Kainz\inst{1,6}\orcidID{0000-0002-7813-5023}
}
\authorrunning{S. Budd et al}
\institute{Imperial College London, Dept. Computing, BioMedIA, London, UK \and
King’s College London, London, UK \and
Guy’s and St Thomas’ NHS Foundation Trust, London, UK \and
School of Health Sciences, City, University of London, London, UK \and
Klinikum Rechts der Isar, Technical University of Munich, Munich, DE\and
Friedrich--Alexander University Erlangen--N\"urnberg, DE\\
\email{samuel.budd13@imperial.ac.uk}}
\maketitle              
\begin{abstract}
Fetal ultrasound screening during pregnancy plays a vital role in the early detection of fetal malformations which have potential long-term health impacts. The level of skill required to diagnose such malformations from live ultrasound during examination is high and resources for screening are often limited. We present an interpretable, atlas-learning segmentation method for automatic diagnosis of Hypo-plastic Left Heart Syndrome (HLHS) from a single `4 Chamber Heart' view image. We propose to extend the recently introduced Image-and-Spatial Transformer Networks (Atlas-ISTN) into a framework that enables sensitising atlas generation to disease. In this framework we can jointly learn image segmentation, registration, atlas construction and disease prediction while providing a maximum level of clinical interpretability compared to direct image classification methods. As a result our segmentation allows diagnoses competitive with expert-derived manual diagnosis and yields an AUC-ROC of 0.978 (1043 cases for training, 260 for validation and 325 for testing). 
\keywords{Segmentation \and Atlas \and Ultrasound}
\end{abstract}

\section{Introduction}

Fetal Ultrasound (US) screening is a key part of ensuring the ongoing health of fetuses during pregnancy. Assessment of fetal development and accurate anomaly detection from US scans are integral in diagnosing potential fetal development issues at the earliest time possible to ensure the best care may be given. For these reasons a mid-trimester US scan is carried out between 18-22 weeks gestation in many countries as part of standard prenatal care procedures. During screening 'standard plane' views are used to acquire images in which key anatomical features may be examined, biometrics extracted and diagnosis of developmental issues may be made \cite{Fasp2018}. Several of these standard views and surrounding frames are used to make the diagnosis of Hypo-plastic Left Heart Syndrome (HLHS). Antenatal diagnosis of congenital heart disease such as HLHS has been shown to result in reduced mortality and morbidity of affected infants~\cite{holland2015prenatal,calderon2012impact}. Unfortunately, antenatal detection of HLHS is not universal, due to the high level of skill required to make the diagnosis accurately from often noisy and inconsistent US views, which vary with gestational age, among other factors.

Recently, automatic ultrasound US scanning methods have been developed using deep learning, mitigating the difficulties of manual US screening through automatic detection of diagnostically relevant anatomical planes~\cite{baumgartner2017sononet}. These systems have enabled the development of robust automated methods for estimation of anatomical biometrics and diagnosis of fetal structural malformations such as HLHS, under diverse acquisition scenarios with various imaging artefacts. Critically, these methods still provide limited interpretability of predictions, and as such reasoning about diagnosis and appropriate interventions remains a challenge even in the presence of accurate predictions of anatomical features and diagnosis of development issues~\cite{Tan2020AutomatedScreening,Miceli2015AAnomalies,Arnaout2018Deep-learningLesions,Yeo2013FetalHeart}.

\noindent\textbf{Related work:}
Automated segmentation of anatomical structures in US images has been the topic of significant research, with CNN based methods~\cite{Budd2019ConfidentSonographers,Sinclair2018Human-levelNetworks,Wu2017CascadedSegmentation} often outperforming non-deep learning approaches~\cite{Carneiro2008DetectionTree,Li2018AutomaticFitting,Rueda2014EvaluationChallenge}.
Many of these methods perform well despite having no prior knowledge of the anatomical structure under consideration. However, in cases where performance drops, the resulting segmentations often bear no resemblance to the expected anatomical structure, resulting in segmentations that are not suitable for downstream analysis. As such, recent work to mitigate this fact has been introduced. 

Methods such as Stochastic Segmentation Networks (SSNs)~\cite{Monteiro2020StochasticUncertainty} aim to enforce continuity between anatomical structure segmentations (an assumption that holds in our case) to force a prediction to segment structures such that they remain connected and allow for sampling multiple plausible solutions to any given image segmentation. Similarly, \cite{clough2019explicit,Hu2019} introduces topological priors to enforce continuity between segmented regions. Another recent approach aims to automatically learn an atlas of the anatomical structure under consideration during training of a segmentation model. Predicting both an image segmentation and a transformation between the automatically constructed atlas and the predicted segmentation, forces the resulting segmentation to retain the expected anatomical structure. In the presence of imaging artefacts or other image features the above behavior may result in a worse segmentation performance~\cite{Sinclair2020Atlas-ISTN:Networks}. The aforementioned methods provide accurate segmentations of anatomical structures familiar to sonographers, however at present, these are not used to perform diagnosis of CHD or provide any means for disease-specific conditioning.

Deep Ultrasound Classification is currently the only option that has been explored in literature to perform automated diagnosis of CHD directly from US images. Deep classification methods achieve high accuracy~\cite{Sushma2021ClassificationCHD,Arnaout2020Expert-levelLearning,Tan2020AutomatedScreening}, but rely on large curated datasets~\cite{Arnaout2020Expert-levelLearning} or additional views of the heart to support multitask learning~\cite{Tan2020AutomatedScreening}. Unfortunately, conventional image classification is difficult to apply to fetal ultrasound because only very specific ``standard planes'' contain sufficient diagnostic information~\cite{Fasp2018}. 
Classifying non-diagnostic frames (that should not be considered healthy or diseased) could lead to erroneous diagnosis or obscure the signal from the true diagnostic frames. As such, direct classification models have very little utility if clinicians cannot clearly interpret and assess the validity of the classification or find a view in pathological cases that would correspond to the defined anatomical standard~\cite{Fasp2018}. 
\noindent\textbf{Contribution:}
In this paper we introduce a novel method for the diagnosis of HLHS from US images using pathology-robust segmentation. To the best of our knowledge, we present for the first time a  segmentation network that is able to jointly segment, register and build a labeled atlas that focuses on relevant features to robustly diagnose HLHS for fetal 4-chamber views in ultrasound imaging. 
By extending the recently proposed Atlas-ISTN framework~\cite{Sinclair2020Atlas-ISTN:Networks} with an additional classification module and corresponding component in the loss function, our method provides an interpretable and accurate option for HLHS diagnosis compared to direct image classification approaches through segmentation of anatomical structures known to sonographers.

We assess the quality of our segmentations in the downstream task of inferring HLHS status from ventricular areas.
From ground truth expert annotations we evaluate the possible correlation of this approximation to true disease status, which is confidently known from post-natal outcome records. We compare this correlation to using naive segmentation for area parameter extraction and our proposed method.  

\section{Method}

\begin{figure}[ht]
     \centering
     \begin{subfigure}[b]{0.24\textwidth}
         \centering
         \includegraphics[width=\textwidth]{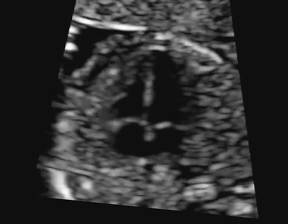}
         \caption{}
     \end{subfigure}
     \hfill
     \begin{subfigure}[b]{0.24\textwidth}
         \centering
         \includegraphics[width=\textwidth]{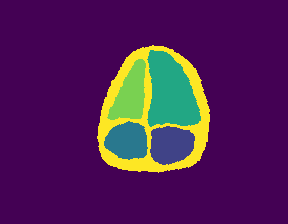}
         \caption{}
     \end{subfigure}
     \hfill
     \begin{subfigure}[b]{0.24\textwidth}
         \centering
         \includegraphics[width=\textwidth]{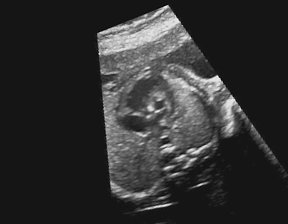}
         \caption{}
     \end{subfigure}
     \hfill
     \begin{subfigure}[b]{0.24\textwidth}
         \centering
         \includegraphics[width=\textwidth]{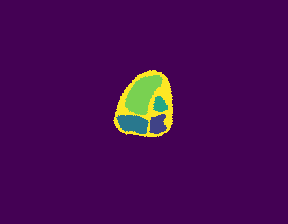}
         \caption{}
     \end{subfigure}
        \caption{Example ultrasound images and manual segmentations of anatomical areas. (a) Healthy patient 4CH ultrasound view; (b) manual segmentation of anatomical areas of healthy heart in (a); (c) HLHS patient's  approximation of a 4CH ultrasound view; (d) manual segmentation of anatomical areas in (c).}
        \label{fig:examples}
\end{figure}

We propose a new method for the automatic diagnosis of HLHS from a single US image of the `4-Chamber Heart View' (4CH view). Our system is inspired by current clinical practice and can be broken down into three major modules. First, for a given 4CH image, we seek a model that can provide accurate segmentations for 5 anatomical areas: \textit{`Whole Heart'}, \textit{`Left Ventricle'}, \textit{`Right Ventricle'}, \textit{`Left Atrium'} and \textit{`Right Atrium'}. Figure~\ref{fig:examples} shows an example for the differences in these areas between a healthy fetus and a baby with HLHS.

\begin{figure}
    \centering
    \includegraphics[width=\textwidth]{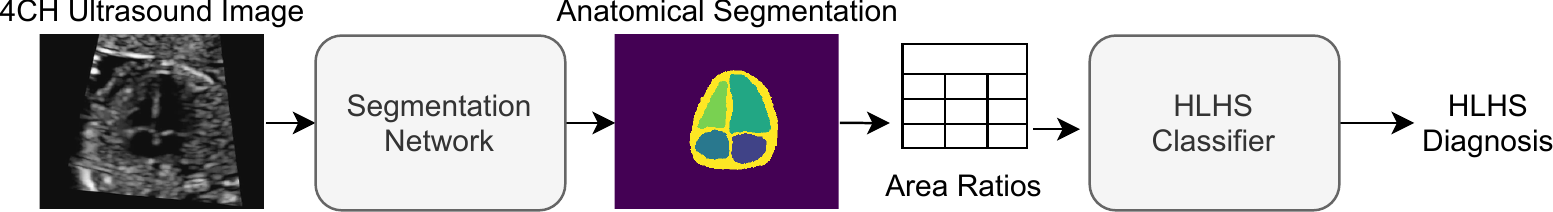}
    \caption{Classification via deep segmentation and quantitative area ratio feature extraction}
    \label{fig:overview}
\end{figure}

Secondly, the resulting segmentation is used to extract simple image features informative of HLHS diagnosis. For each class, we calculate the ratio of that region's area to the area of every other segmented region to obtain a set of scalar features representative of that image. 
Finally, we use the quantitative features to construct a classifier for HLHS using: 1) class-weighted Logistic Regression Classifier as baseline; 2) Gaussian Process classifier with a radial basis basis function kernel. Figure~\ref{fig:overview} shows an overview over the diagnostic approach. 

For the task of \textbf{Robust segmentation}, we adopt a recently proposed method, Atlas-ISTN~\cite{Sinclair2020Atlas-ISTN:Networks}, that generates a segmentation as well as learns a label atlas that both ensures robustness and can be used to inspect the inner beliefs of the network. Conditioning is used to sensitise the atlas generation to regions that are most relevant for the downstream task of disease classification. 

\begin{figure}
    \centering
    \includegraphics[width=0.9\textwidth]{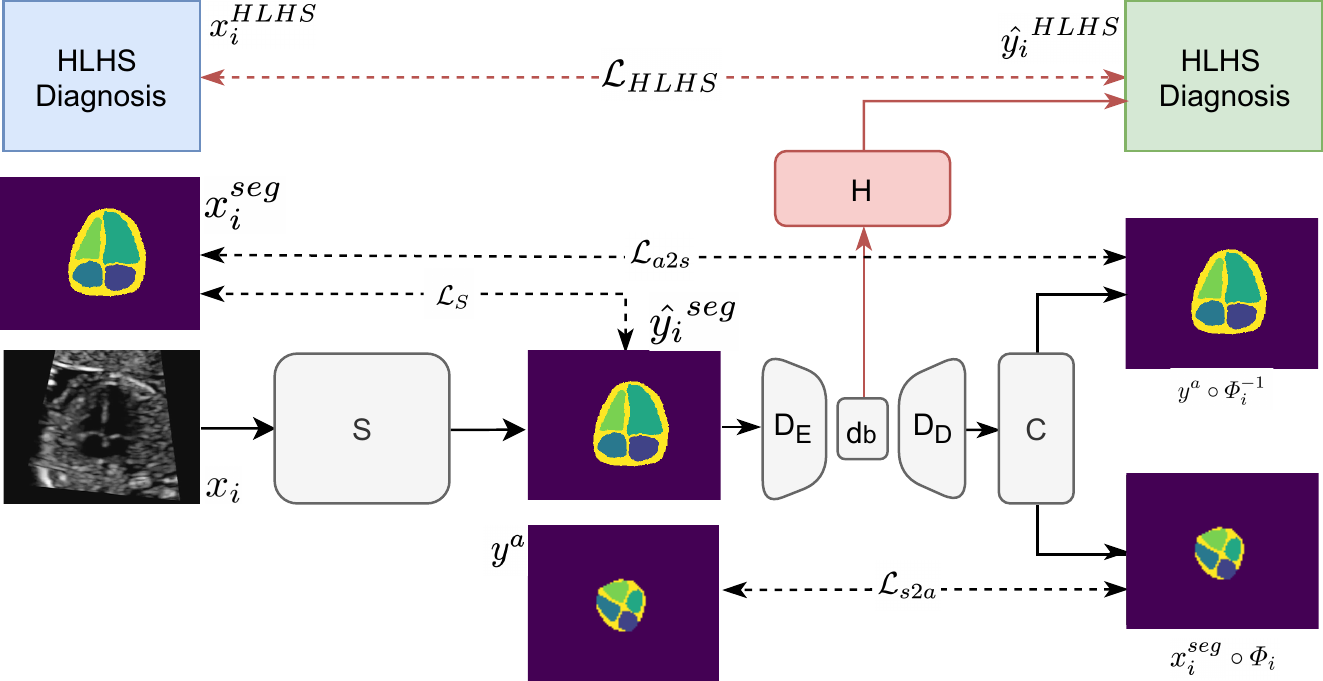}
    \caption{Disease conditioned Atlas-ISTN network architecture.
    $\textbf{S}$ the segmentation network;
    $\textbf{D}$ indicates the atlas to image mapping module;
    $\textbf{C}$ is the transformation computation Module; 
    and 
    $\textbf{H}$ is a disease prediction branch, highlighted in red, which is the key difference to~\cite{Sinclair2020Atlas-ISTN:Networks}.}
    \label{fig:cistn}
\end{figure}

Our detailed model is outlined in Figure~\ref{fig:cistn}. As input we use uncropped 4-chamber ultrasound images, ground truth segmentation maps and a binary disease label in 
$
    X = \{x_i, x_i^{seg},x_i^{HLHS}\}.
$
The model aims to learn: 
\begin{align*}
    \{\hat{y_i}^{seg},\hat{y_i}^{HLHS}, y^a\} = \textbf{M}
    (x_i),
\end{align*}
where \textbf{M} 
is the entire model, $\hat{y_i}^{seg}$ are the logits of a predicted segmentation  describing five cardiac labels with one background channel as defined in ${x_i}^{seg}$,  
$\hat{y_i}^{HLHS}$ are probabilities for discrete disease categories in ${x_i}^{HLHS} \in [0,1]$, and $y^a$ is an automatically optimised  atlas label map.
$\mathbf{M}$ consists of four modules. Image to segmentation mapping is obtained through 
$
\hat{y_i}^{seg} = 
\mathbf{S}_{\theta_s}(x_i),
$
which we define as a 2D UNet~\cite{RonnebergerU-Net:Segmentation} with a SSN module~\cite{Monteiro2020StochasticUncertainty}. 
The concatenation of $\hat{y_i}^{seg}$ and the atlas label map ${y}^{a}$ is used to establish the atlas to image transformation:
\begin{align*}
\mathbf{d}_{b_i} = \mathbf{D}_{E,\theta_{enc}}(\hat{y_i}^{seg},{y}^{a}), \\
\{v_i,T_i\} = \mathbf{D}_{D,\theta_{dec}}(\mathbf{d}_{b_i}).  
\end{align*}
$v_i$ is a stationary velocity field and $T_i$ an affine transformation matrix that is processed to a deformation field with a Transformation Computation Module $\mathbf{C}$ according to~\cite{Sinclair2020Atlas-ISTN:Networks}. Thus, $\mathbf{C}$ yields forward and inverse transformations, $\Phi_i$ and $\Phi^{-1}_i$.
To steer the atlas generation process and emphasise disease-relevant labels, we predict $\hat{y_i}^{HLHS}= \mathbf{H_{\theta_h}}(\mathbf{d}_{b_i})$, where $\mathbf{H}_{\theta_h}$ are three fully connected layers with ReLU activations. 

Additionally to the image transformer loss:
\begin{align*}
    \mathcal{L}_{S} &= \frac{1}{N}  \left( \sum_{i=1}^{N} ||{x_i}^{seg} - \hat{y_i}^{seg} ||^2 \right ),
\end{align*}
the atlas-to-segmentation loss: 
\begin{align*}
    \mathcal{L}_{a2s} = \frac{1}{N}  \left( \sum_{i=1}^{N} \sum_{j=1}^{c} ||{x_{i,j}}^{seg} - {y_j}^{a} \circ \Phi^{-1}_i ||^2 \right),
\end{align*}
and the segmentation-to-atlas loss:
\begin{align*}
    \mathcal{L}_{s2a} =  \frac{1}{N}  \left( \sum_{i=1}^{N} \sum_{j=1}^{c} ||{x_{i,j}}^{seg} \circ \Phi_i -  {y_j}^{a}  ||^2 \right),
\end{align*}
where $j$ indicates the individual labels,  we introduce a cross entropy loss term:
\begin{align*}
    \mathcal{L}_{HLHS} = - ({x}_{i}^{HLHS} \log(\hat{y_i}^{HLHS}) + (1 - {x}_{i}^{HLHS})  \log(1 - \hat{y}_{i}^{HLHS}))
\end{align*}
to enforce disease-sensitive atlas generation. Thus, our final loss function is the Atlas-ISTN loss~\cite{Sinclair2020Atlas-ISTN:Networks} with its regularisation loss term, $\mathcal{L}_{reg} = \sum_{i}^{N} ||\nabla \phi||^2$, that encourages smoothness of the non-rigid deformation fields, paired with $\mathcal{L}_{HLHS}$:
\begin{align*}
    \mathcal{L} =\mathcal{L}_{S} +\omega(\mathcal{L}_{a2s} + \mathcal{L}_{s2a} + \lambda \mathcal{L}_{reg}) + \gamma \mathcal{L}_{HLHS},
\end{align*}
where $\lambda$ adjusts smoothness of $\Phi$, $\omega$ influences the contribution of the deformation terms similar to \cite{Sinclair2020Atlas-ISTN:Networks}, and $\gamma$ steers how much the atlas should be specific to the targeted disease category. 

\noindent\textbf{HLHS classification from fetal cardiac 4-chamber view segmentations:} 
We extract numerical features from $\hat{y_i}^{seg}$ in order to classify HLHS vs. Healthy patients from interpretable features $f = \{f_0, f_1, ..., f_N\}$ where $f_i = r_{ab} = A_a / A_b$ if $a \neq b$ and $r_{ba}$ is not in $f$ already. We represent the ratio between two quantities as $r_{ab}$ and consider $r_{ab}$ and $r_{ba}$ to contain equivalent information and as such exclude the latter from $f$. Here $A_a$ is the count of pixels belonging to class $a$ in $\hat{y_i}^{seg}$ which acts as an estimate to the area.

We apply two common classification algorithms to classify the extracted segmentation area ratio features as healthy vs HLHS. We first use an L2 regularised, class weight balanced Logistic regression classifier implementation. Secondly we use a Gaussian Process classifier based on Laplace approximation~\cite{GPML}.

\section{Experiments and Results}

\noindent\textbf{Data and Pre-processing}:
We use a private and Ethics/IP-restricted, de-identified dataset of 1628 4CH US images (1560 healthy controls, 68 HLHS), with 1043 for training, 260 for validation, 325 for testing with equivalent class imbalance within each set (42, 10 and 16 HLHS cases respectively), acquired on Toshiba Aplio  i700, i800 and Philips EPIQ V7 G devices. Class imbalance reflects the prevalence of HLHS observed in our tertiary care referral clinic ($\sim3-4\%$), which is a specialised centre, thus the incident rate is relatively high. HLHS is rare, $\sim 3$ in $10000$ live births ~\cite{EUprevalence2021}, thus this condition can be challenging to identify for primary care sonographers.
Our images are taken from volunteers at 18-24 weeks gestation (Ethics: \emph{[anonymous during review]}), acquired in a fetal cardiology clinic, where patients are given advanced screening due to their family history. Each image has been hand-picked from ultrasound videos by an expert sonographer, representing a best possible 4CH view. A fetal cardiologist and three expert sonographers delineated the images using Labelbox~\cite{Labelbox}. The images have been resampled to $288 \times 224$ pixels, centred on the heart and aligned along the cardiac main axis.

\noindent\textbf{Robust segmentation}
We compare several methods for automated segmentation by average DICE score achieved for each anatomical class and summarise the results in Table~\ref{tab:allresults}. We show that each of the compared methods is effective for the segmentation of anatomical structures from ultrasound image views. The question remains, which method produces the most informative segmentation for downstream disease diagnostics?

\begin{figure}
    \centering
    \begin{subfigure}[b]{0.95\textwidth}
    \begin{subfigure}[b]{0.15\textwidth}
         \centering
         \includegraphics[width=\textwidth]{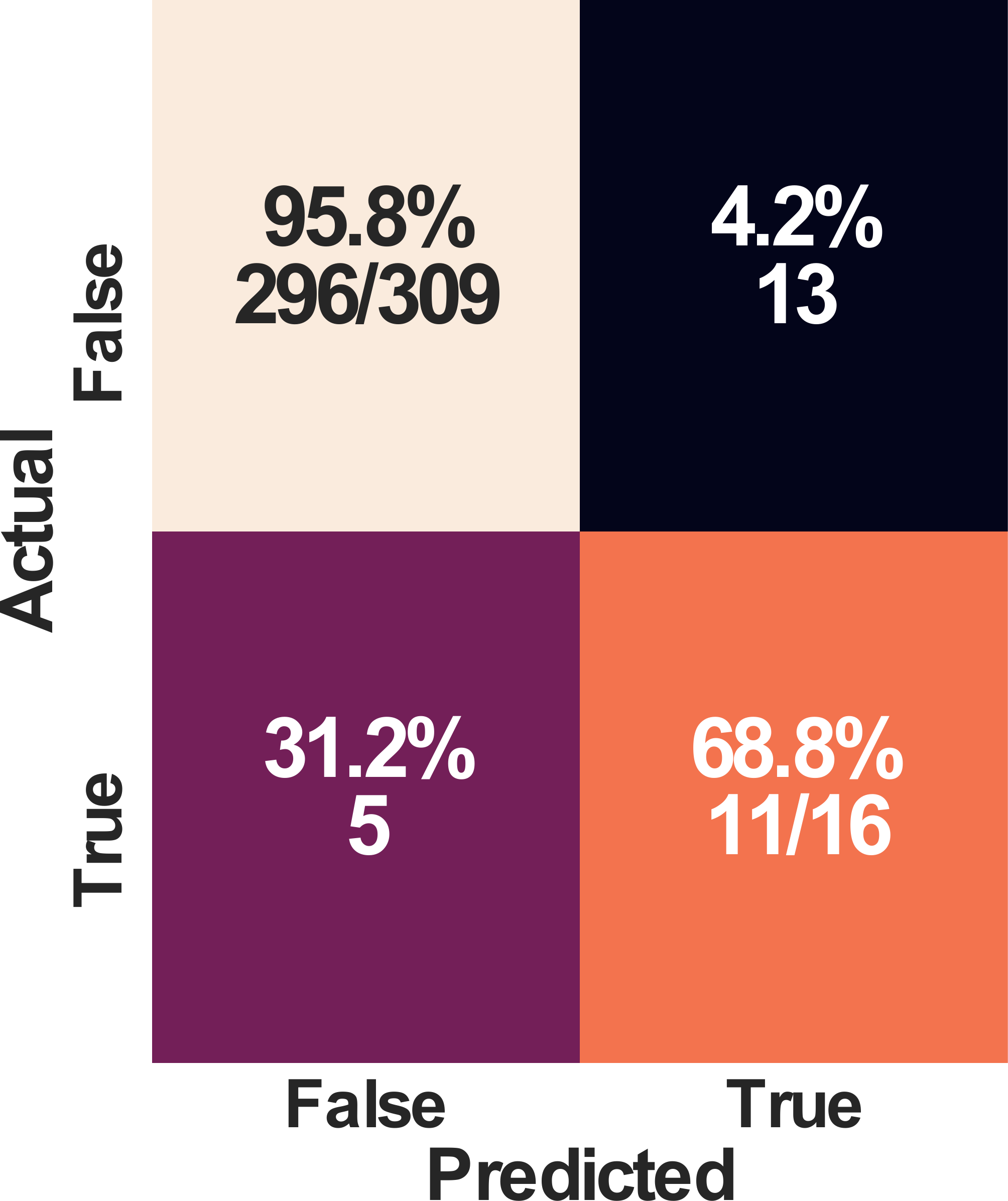}
         \caption{Expert}\label{a}
     \end{subfigure}
     \hfill
     \begin{subfigure}[b]{0.15\textwidth}
         \centering
         \includegraphics[width=\textwidth]{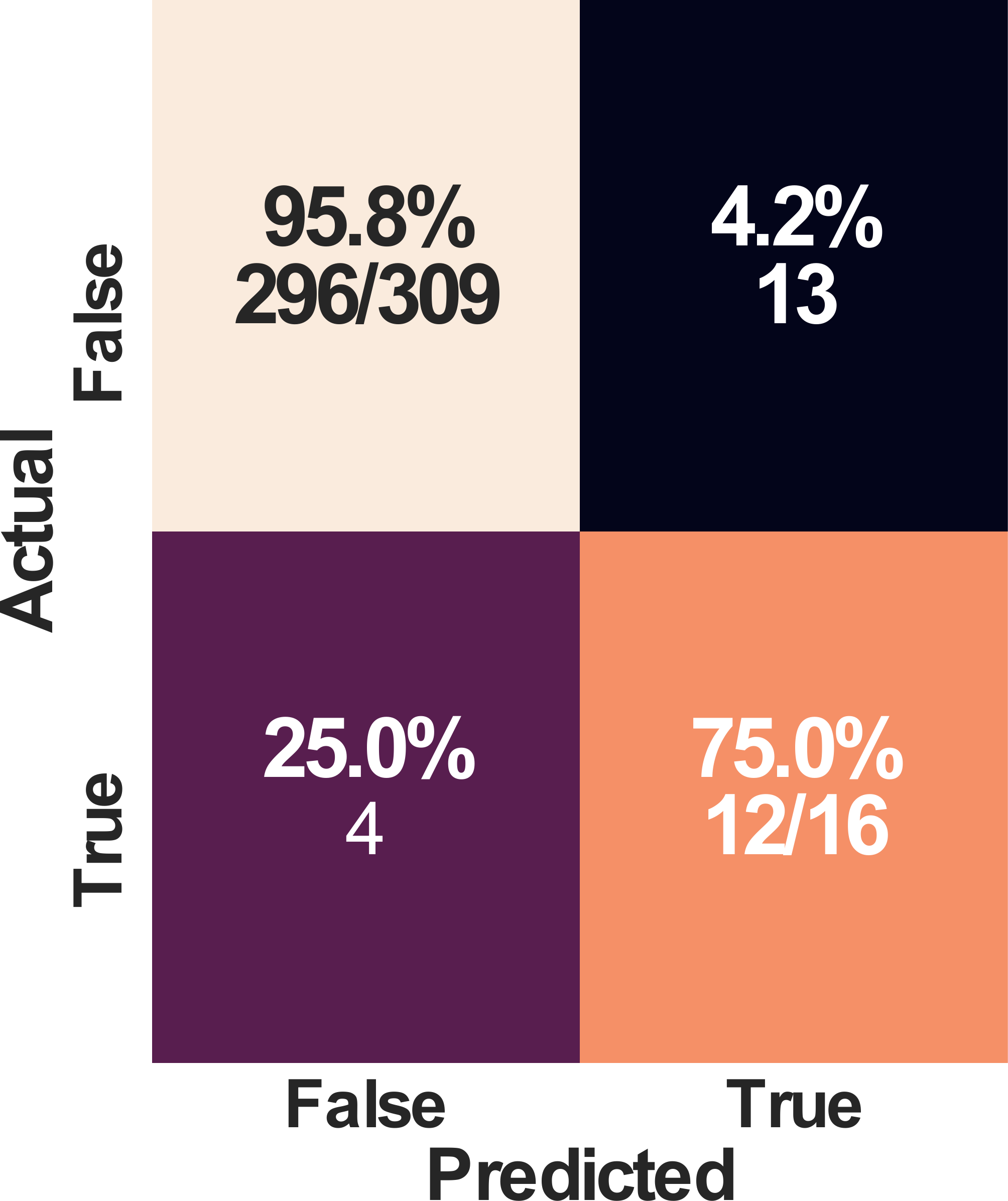}
         \caption{UNet}\label{b}
     \end{subfigure}
     \hfill
    \begin{subfigure}[b]{0.15\textwidth}
         \centering
         \includegraphics[width=\textwidth]{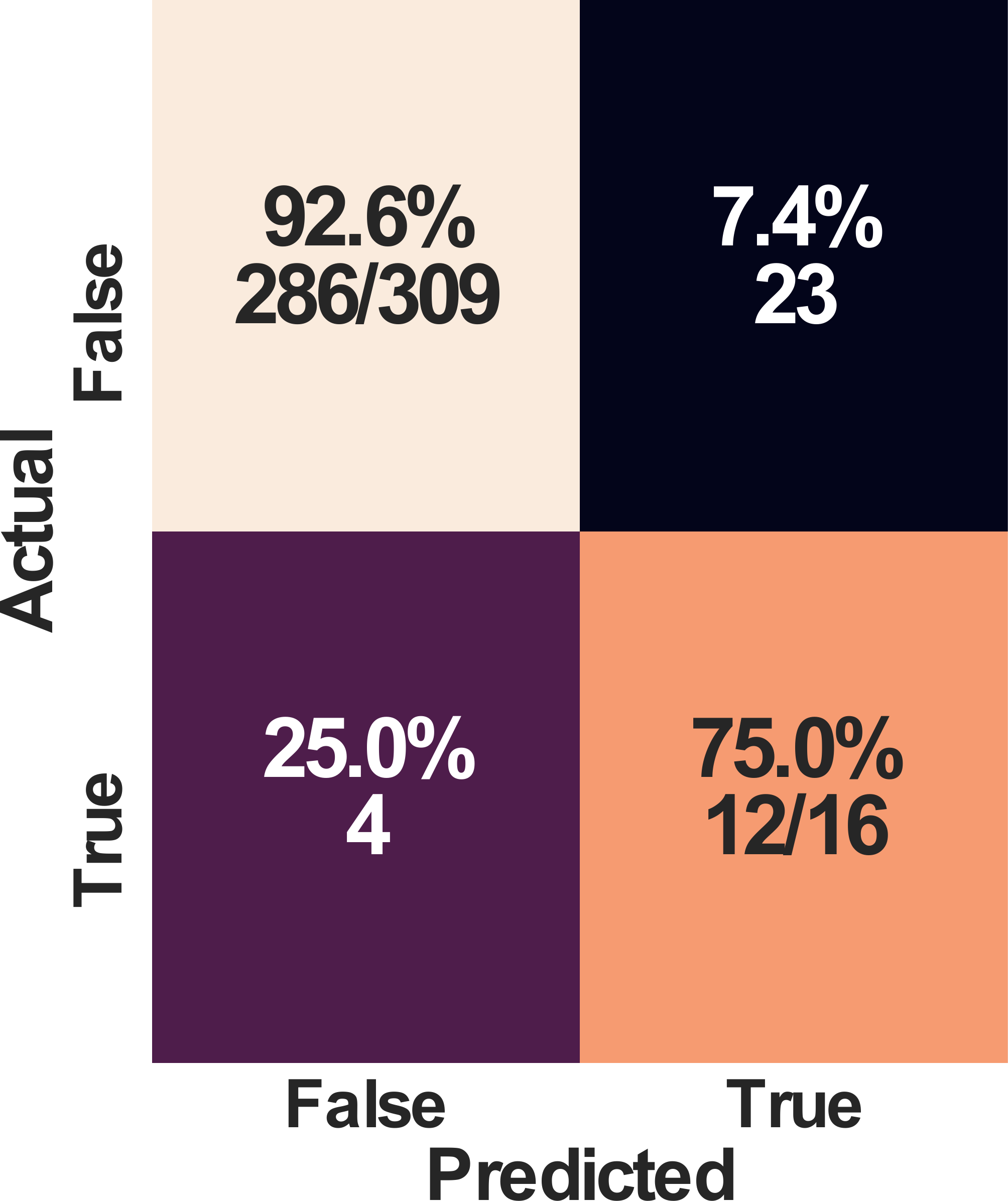}
         \caption{SSN}\label{c}
     \end{subfigure}
     \hfill
     \begin{subfigure}[b]{0.15\textwidth}
         \centering
         \includegraphics[width=\textwidth]{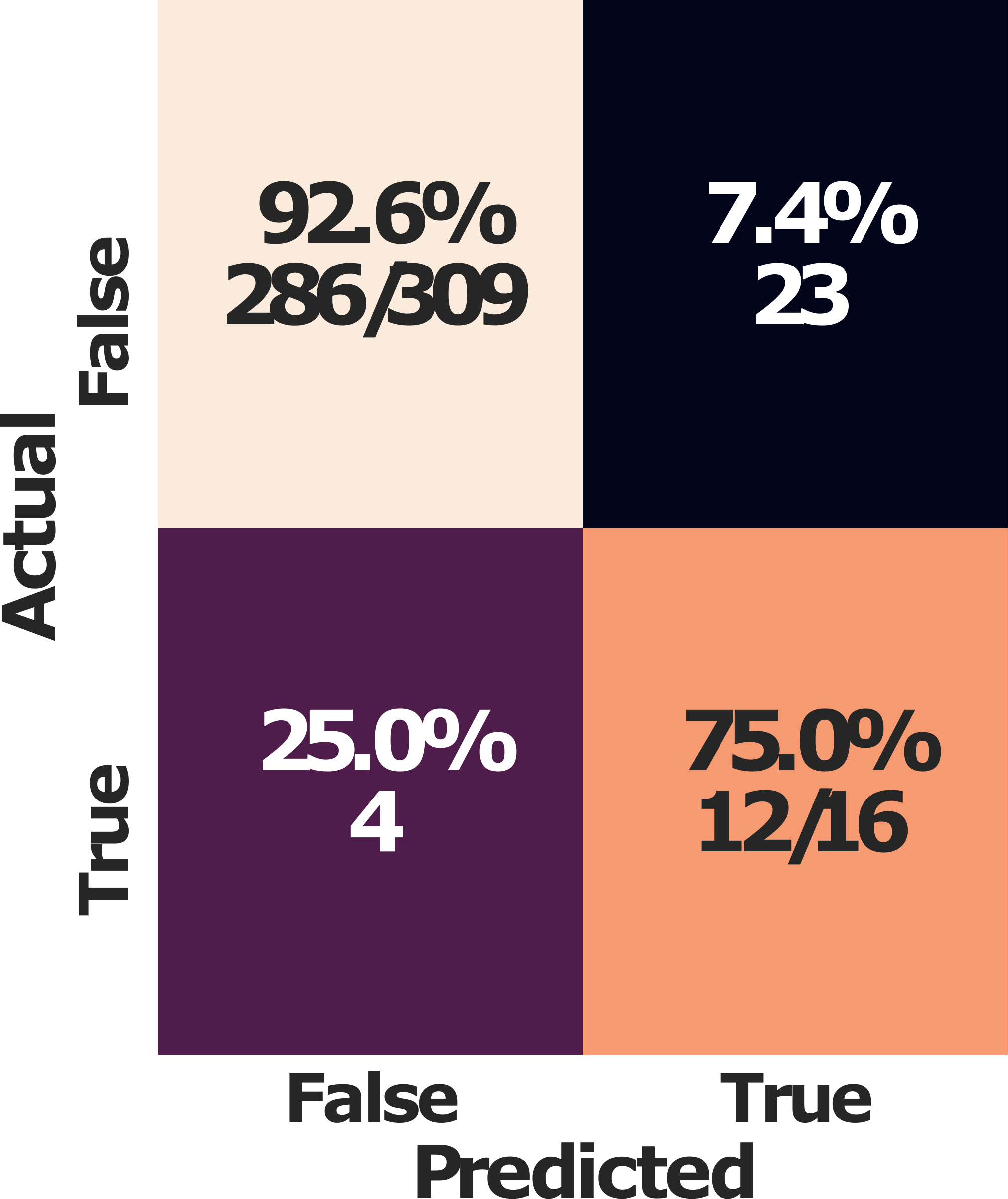}
         \caption{$\gamma=0$}\label{d}
     \end{subfigure}
     \hfill
    \begin{subfigure}[b]{0.15\textwidth}
         \centering
         \includegraphics[width=\textwidth]{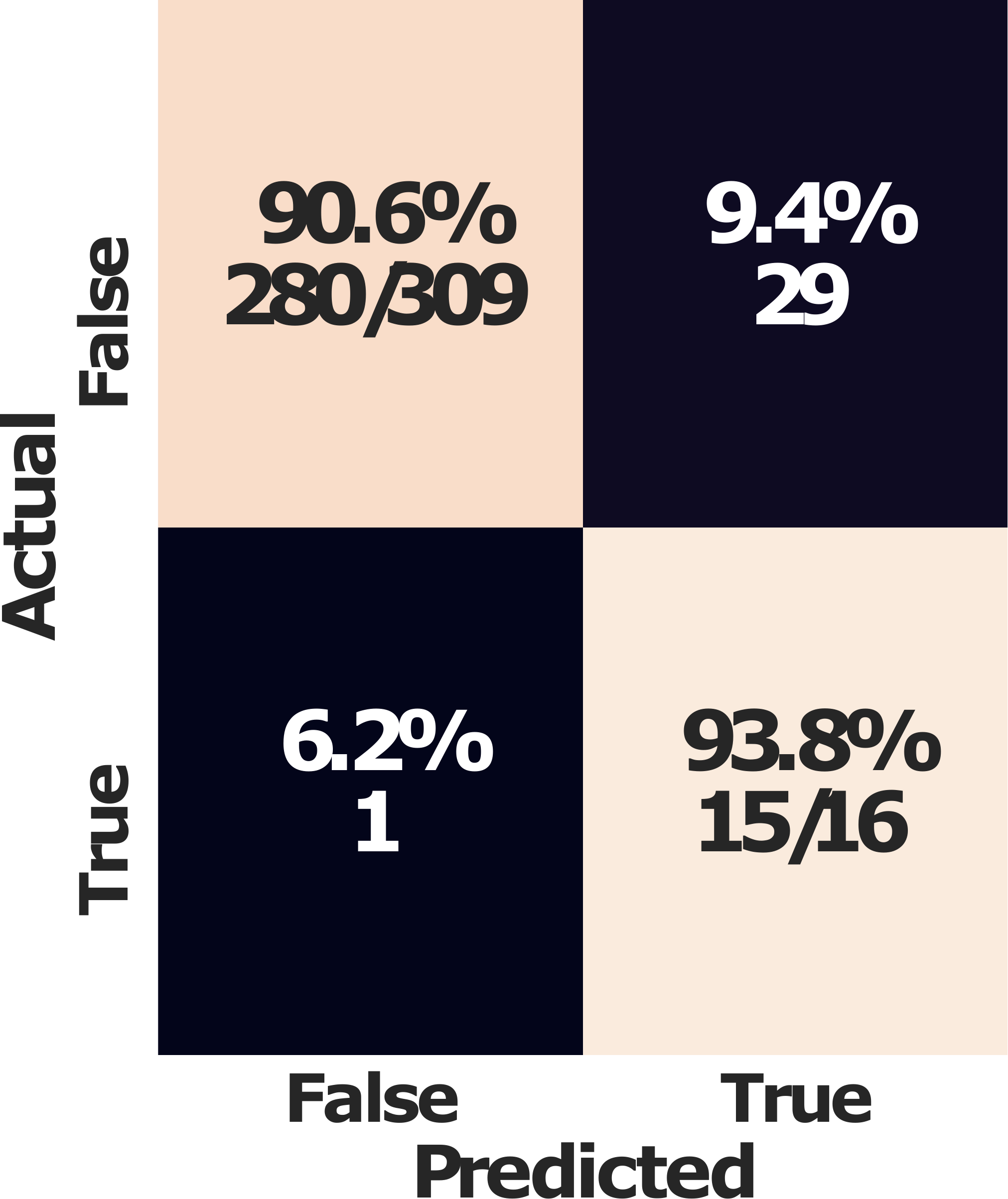}
         \caption{$\lambda=1$}\label{e}
     \end{subfigure}
     \hfill
     \begin{subfigure}[b]{0.15\textwidth}
         \centering
        \includegraphics[width=\textwidth]{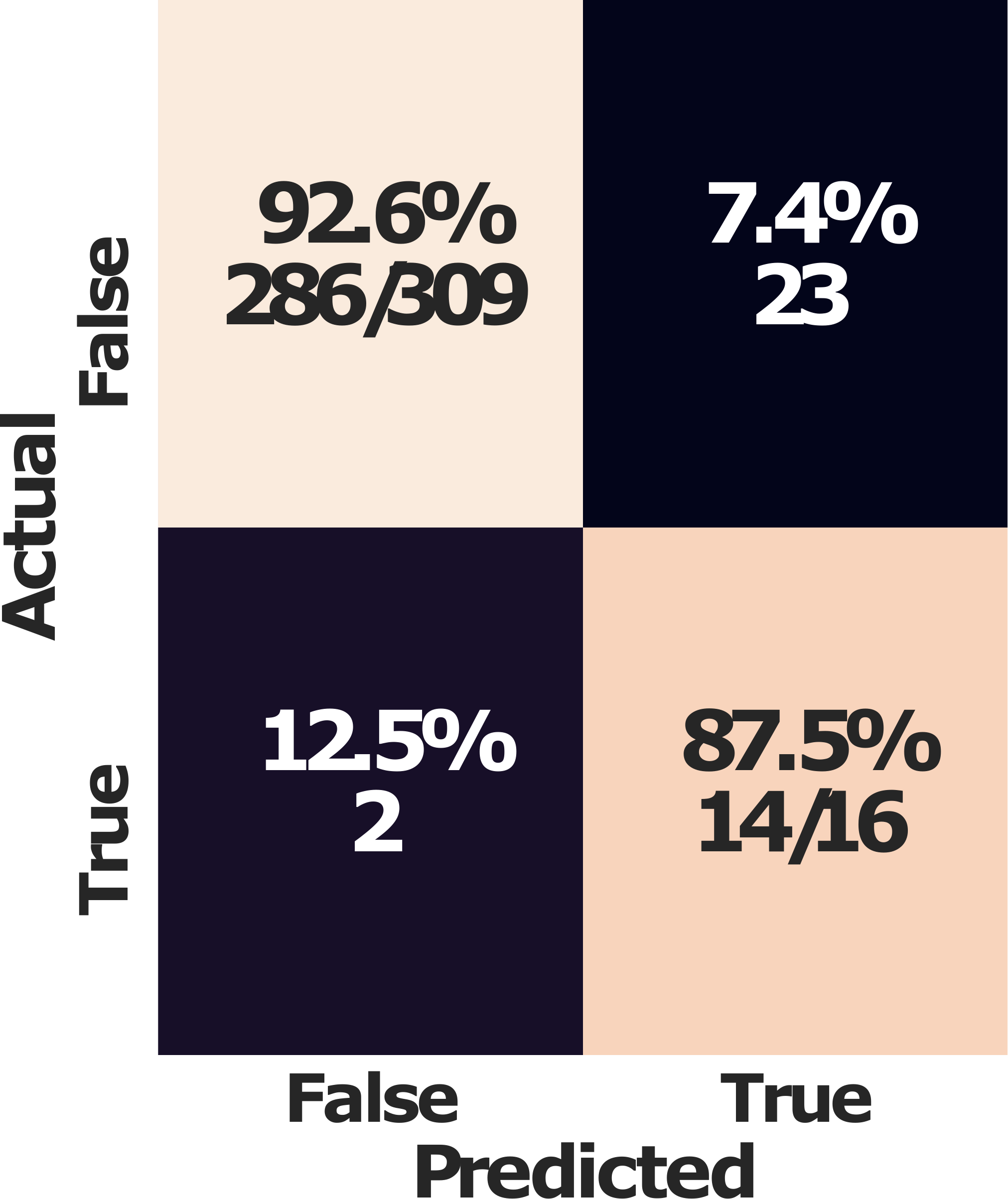}
         \caption{$\lambda=1000$}\label{f}
     \end{subfigure}
     \vfill
     \begin{subfigure}[b]{0.15\textwidth}
         \centering
         \includegraphics[width=\textwidth]{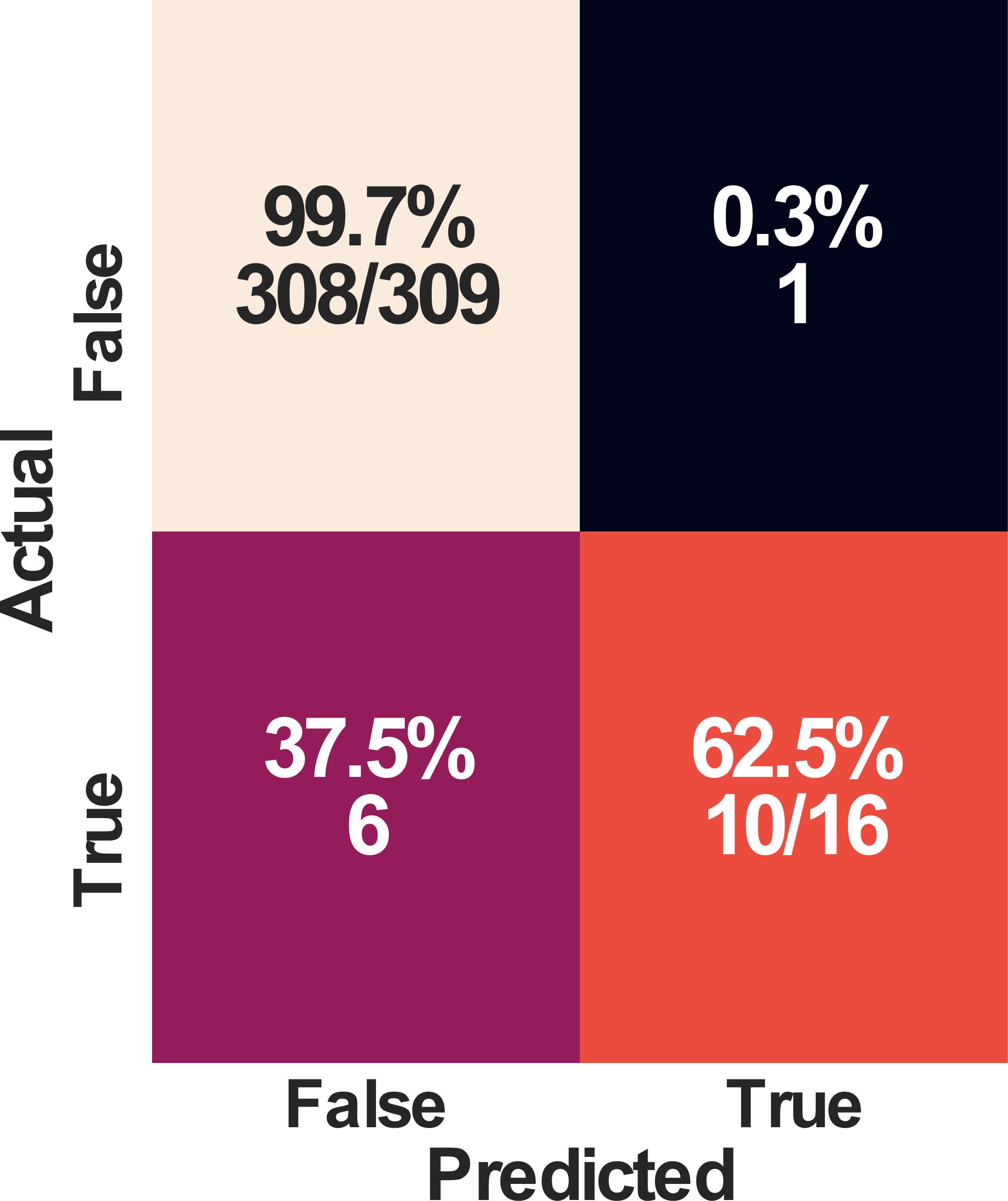}
         \caption{Expert}\label{g}
     \end{subfigure}
     \hfill
     \begin{subfigure}[b]{0.15\textwidth}
         \centering
         \includegraphics[width=\textwidth]{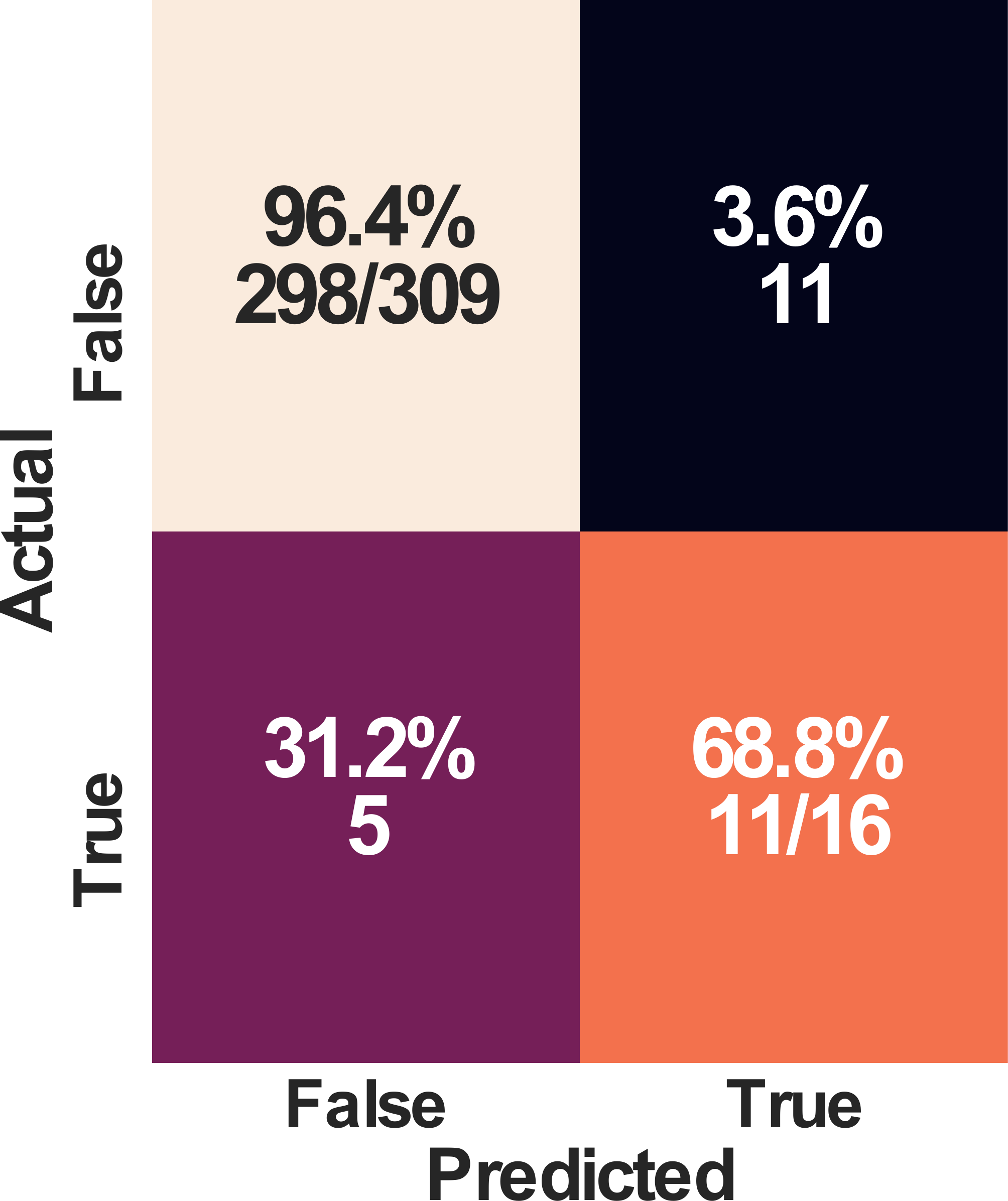}
         \caption{UNet}\label{h}
     \end{subfigure}
     \hfill
    \begin{subfigure}[b]{0.15\textwidth}
         \centering
         \includegraphics[width=\textwidth]{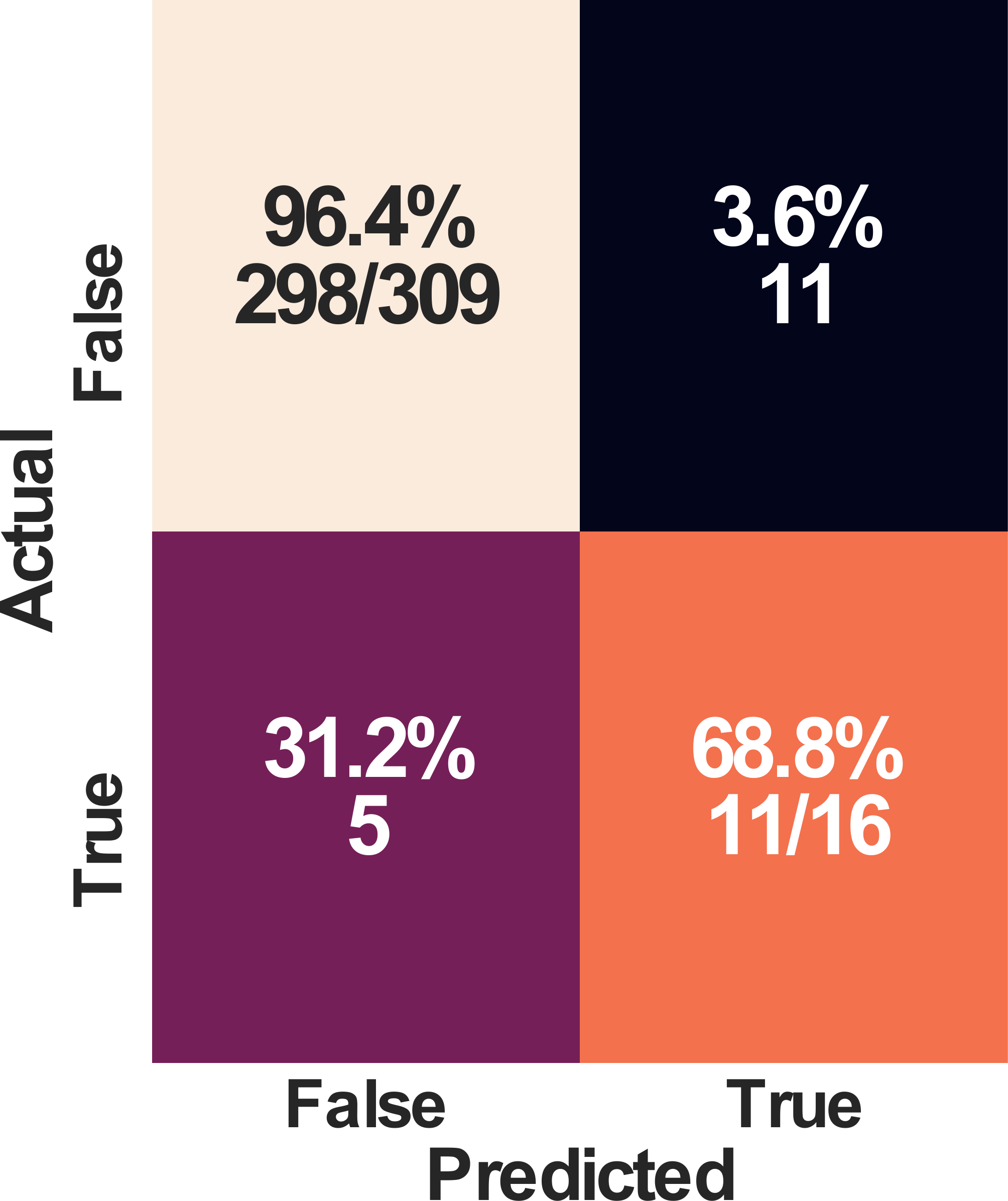}
         \caption{SSN}\label{i}
     \end{subfigure}
     \hfill
     \begin{subfigure}[b]{0.15\textwidth}
         \centering
         \includegraphics[width=\textwidth]{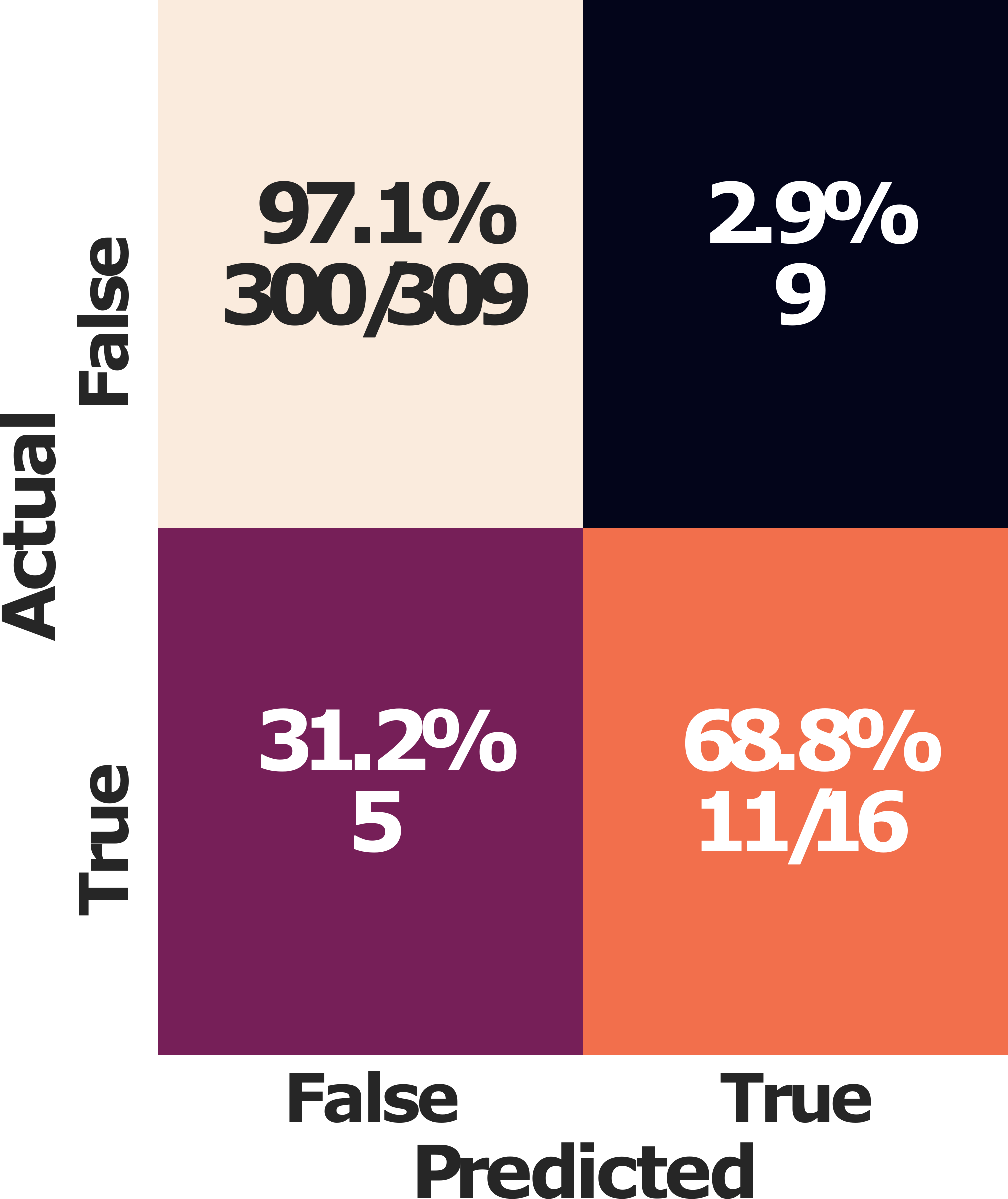}
         \caption{$\gamma=0$}\label{j}
     \end{subfigure}
     \hfill
    \begin{subfigure}[b]{0.15\textwidth}
         \centering
         \includegraphics[width=\textwidth]{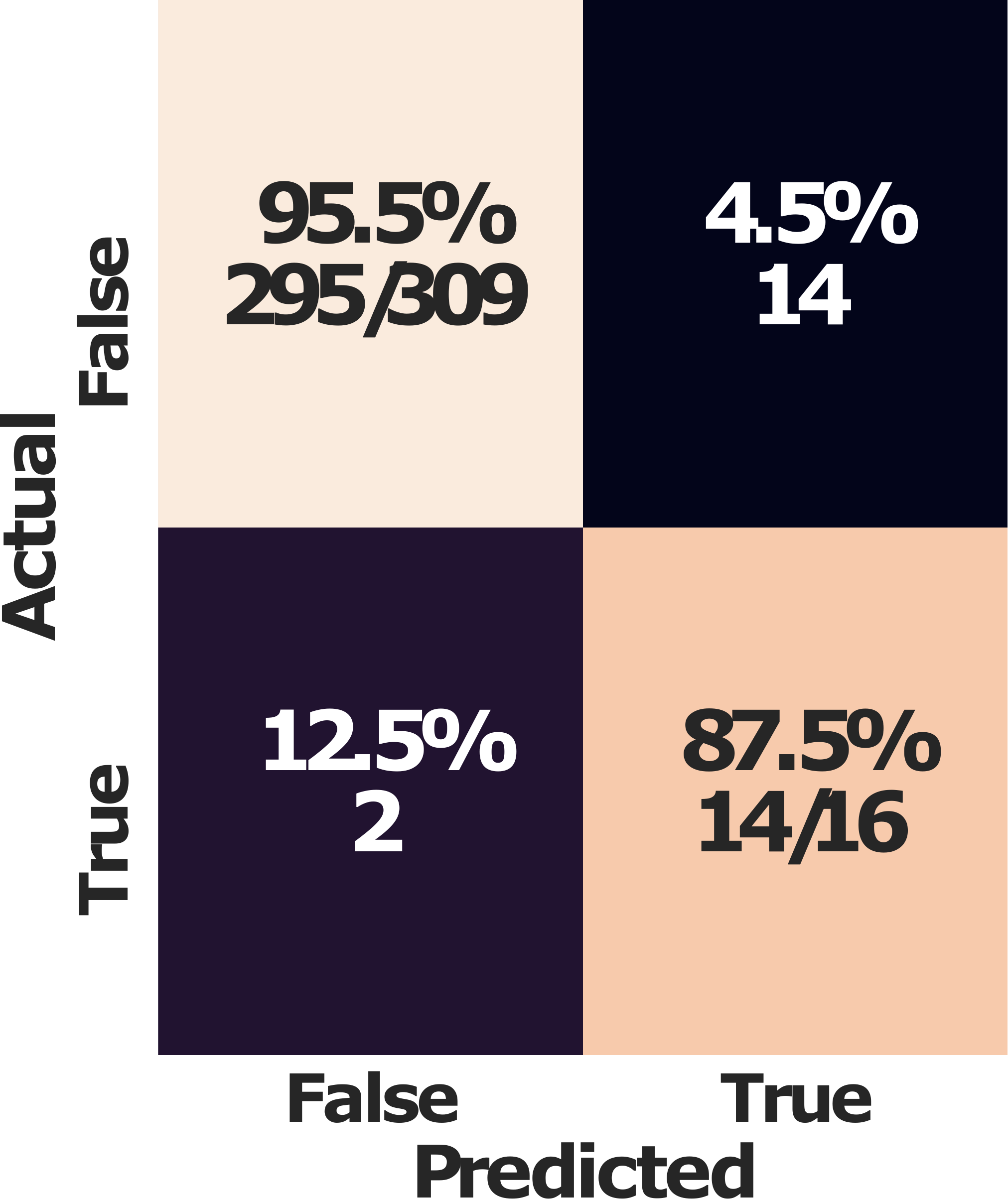}
         \caption{$\lambda=1$}\label{k}
     \end{subfigure}
     \hfill
     \begin{subfigure}[b]{0.15\textwidth}
         \centering
         \includegraphics[width=\textwidth]{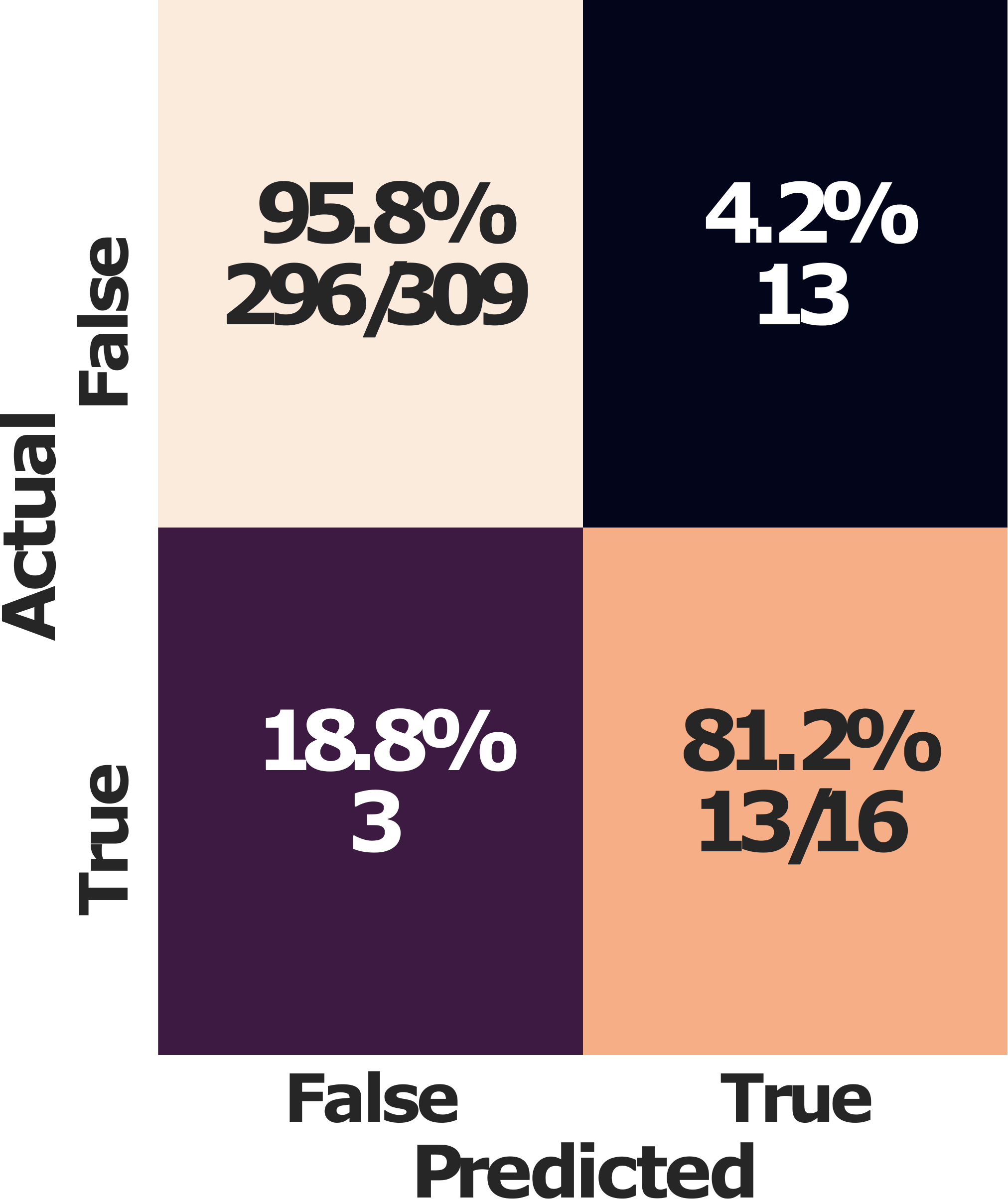}
         \caption{$\lambda=1000$}\label{l}
     \end{subfigure}
     \end{subfigure}
     \begin{subfigure}[b]{0.04\textwidth}
      \raisebox{1.0cm}{\includegraphics[width=\textwidth]{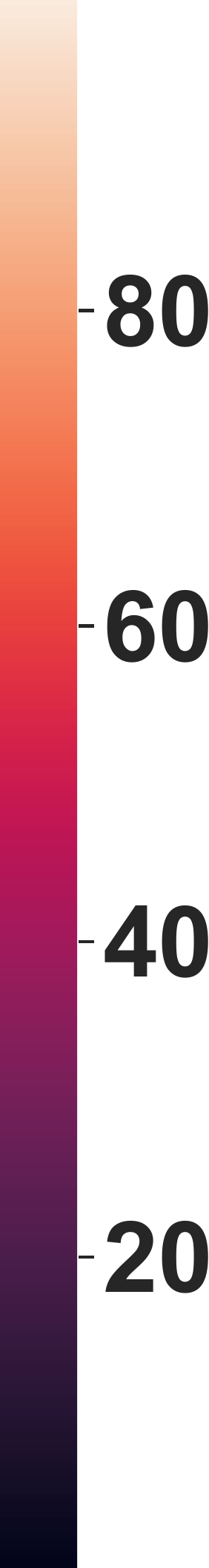}}
     \end{subfigure}
    \caption{Confusion matrices for expert derived classification: Top Row shows logistic regression, bottom row shows Gaussian Process. $\gamma=1$ for (e-f)(k-l).}
    \label{fig:confs}
\end{figure}

\noindent\textbf{Expert derived single image classification:}
To establish human performance on the segmentation task, heart segmentation features (area ratios of each anatomical class) are extracted from the manual ground truth segmentations. These are used to train a linear classifier and a Gaussian process classifier to predict HLHS diagnosis. We report the confusion matrices for each method using the ground truth segmentations shown in Figure~\ref{fig:confs}. Table~\ref{tab:allresults} reports F1-score for positive and negative HLHS classification as well as ROC-AUC for manual as well as automated segmentation.

F1 and AUC scores in Table \ref{tab:allresults} show that our ‘Area Ratios’ classification method achieves state-of-the-art performance for HLHS classification over previous classification methods. Classification performance of ‘area ratios’ extracted from automated segmentations is on par with those extracted from expert manual segmentations. The addition of a disease-conditioned branch to the Atlas-ISTN improves the downstream ‘area ratios’ classification task performance over both expert segmentations and previous segmentation methods.

Figure \ref{fig:confs} shows the performance of the ‘area ratios’ classification using segmentations produced by experts and by each tested segmentation method. Subfigures (\ref{e}-\ref{f}) and (\ref{k}-\ref{i}) highlight the improved sensitivity (fewer false negatives) of Atlas-ISTNs with a disease conditioning branch over expert segmentations (\ref{a},\ref{g}) and other segmentation methods (\ref{b}-\ref{d}, \ref{h}-\ref{j}). Our application is for fetal screening and as such sensitivity is the desired metric to improve, and due to the low prevalence of HLHS, F1 scores for HLHS across all methods may seem low.

Table \ref{tab:allresults} shows our diagnostic branch (\textbf{H}) is competitive with previous image classification approaches, further to this our method uses only a single 4CH image as opposed to previous methods that use multiple heart view US images or video sequences. Our method provides greater interpretability by producing a segmentation (from which ‘area ratios’ classification is performed) and a disease specific atlas for free. Examples for constructed atlases with different configuration are shown in Figure~\ref{fig:atlases}. 

\noindent\textbf{Implementation:} PyTorch 1.7.1+cu110 with two Nvidia Titan RTX GPUs used to train segmentation and atlas models ($\sim10^6$ parameters) in 24-48 hours; scikit-learn~\cite{scikit-learn} for the LR and GP models.

\newcommand\hi{1.8cm}
\begin{figure}[ht]
    \centering
    \begin{subfigure}[b]{0.32\textwidth}
         \centering
         \includegraphics[height=\hi,trim=2cm 1cm 2cm 1cm,clip]{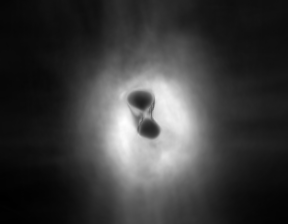}
         \includegraphics[height=\hi,trim=2cm 1cm 2cm 1cm,clip]{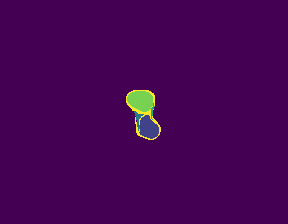}
         \caption{$Atlas^{\gamma=0}_{\lambda=1}$}
     \end{subfigure}
     \hfill
     \begin{subfigure}[b]{0.32\textwidth}
         \centering
         \includegraphics[height=\hi,trim=2cm 1cm 2cm 1cm,clip]{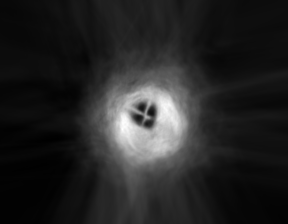}
         \includegraphics[height=\hi,trim=2cm 1cm 2cm 1cm,clip]{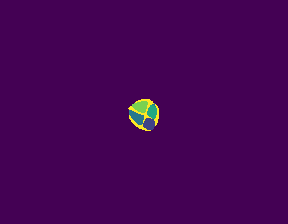}
         \caption{$Atlas^{\gamma=1}_{\lambda=1}$}
     \end{subfigure}
     \hfill
    \begin{subfigure}[b]{0.32\textwidth}
         \centering
         \includegraphics[height=\hi,trim=2cm 1cm 2cm 1cm,clip]{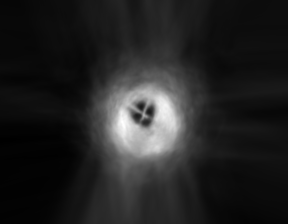}
        \includegraphics[height=\hi,trim=2cm 1cm 2cm 1cm,clip]{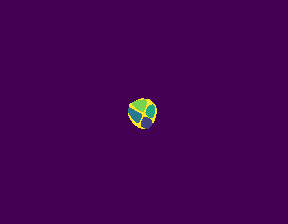}
         \caption{$Atlas^{\gamma=1}_{\lambda=10^3}$}
     \end{subfigure}
    \caption{Example automatically constructed atlas images and atlas label maps in the tested configurations. 
    }
    \label{fig:atlases}
\end{figure}

\begin{table}[t]
\centering
\begin{tabular}{lcccccc|lccc}
\toprule
& \multicolumn{6}{c}{DICE Score} & & \multicolumn{2}{c}{F1 Score} & \multicolumn{1}{c}{ROC}\\
\cmidrule(lr){2-7} \cmidrule(lr){9-10} \cmidrule(lr){11-11}
\textbf{Method}   & \textbf{BG} & \textbf{LA}  & \textbf{RA}  & \textbf{LV}  & \textbf{RV}  & \textbf{WH} & ~~ & \textbf{NC} & \textbf{HLHS} & \textbf{AUC}  \\
\cmidrule(lr){2-7} \cmidrule(lr){9-10} \cmidrule(lr){11-11}
Expert   & 1.000 & 1.000  & 1.000   & 1.000   & 1.000 & 1.000  &  \textbf{LR:} & 0.970 & 0.550 & 0.944 \\
(Std)  & -- & -- & -- & -- & -- & -- &  \textbf{GP} & 0.989 & 0.741 & 0.954\\
UNet~\cite{RonnebergerU-Net:Segmentation}   & 0.993 & 0.768  & 0.804  & 0.793  & 0.794 & 0.635 &  \textbf{LR:} & 0.972 & 0.585 & 0.922 \\
(Std)  & (0.007) & (0.192) & (0.185) & (0.184) & (0.153) & (0.105) &  \textbf{GP:} & 0.974 & 0.579 & 0.928\\
SSN~\cite{Monteiro2020StochasticUncertainty}   & 0.993  & 0.761  & 0.800  & 0.793  & 0.794  & 0.632 &  \textbf{LR:} & 0.955 & 0.471 & 0.883 \\
(Std)  & (0.007) & (0.196) & (0.192) & (0.194) & (0.154 & (0.108) & \textbf{GP:} & 0.974 & 0.579 & 0.923\\
\midrule
$Atlas^{\gamma=0}_{\lambda=1}$  & 0.991  & 0.767  & 0.789  & 0.801  & 0.783  & 0.626 &  \textbf{LR:} & 0.942 & 0.451 & 0.895 \\
(Std)  & (0.007) & (0.187) & (0.192) & (0.191) & (0.172) & (0.106) &  \textbf{GP:} & 0.981 & 0.625 & 0.970\\
$Atlas^{\gamma=1}_{\lambda=10^3}$  & 0.993  & 0.764  & 0.789  & 0.791  & 0.790  & 0.648 &  \textbf{LR:} & 0.958 & 0.528 & 0.929\\
(Std) & (0.007) & (0.185) & (0.184) & (0.196) & (0.146) & (0.110)  &  \textbf{GP:} & 0.974 & 0.619 & 0.973\\
      & -- & -- & -- & -- & -- & -- &  $\mathbf{H:}$ & 0.967 & 0.565 & 0.883  \\
$Atlas^{\gamma=1}_{\lambda=1}$   & 0.993 & 0.760 & 0.783 & 0.784  & 0.788  & 0.637 &  \textbf{LR:} & 0.950 & 0.500 & 0.974\\
(Std) & (0.007) & (0.197) & (0.200) & (0.208) & (0.164) & (0.110) &  \textbf{GP:} & 0.974 & 0.636 & 0.978\\
      & -- & -- & -- & -- & -- & -- &  $\mathbf{H:}$ & 0.982 & 0.667 & 0.905  \\
\bottomrule
\end{tabular}
\caption{DICE scores and standard deviation (Std) for all segmentation methods (left) and performance of downstream disease predictors (right).  (\textbf{BG} = background; \textbf{LA} = left atrium; \textbf{RA} = right atrium; \textbf{LV} = left ventricle; \textbf{RV} = right ventricle; \textbf{WH} = whole heart; \textbf{LR} = Logistic regression; \textbf{GP} = Gaussian process; $\mathbf{H}$ = disease prediction branch; \textbf{NC} = Normal control; \textbf{HLHS} = hypo-plastic left heart syndrome.)
}
\label{tab:allresults}
\end{table}

\section{Discussion}

Assuming a linear downstream model for clinical decision making, our results show that automated segmentation methods are en-par with human-generated annotations for the accurate identification of HLHS patients. We believe the reason for higher than expert performance is due to more consistent automated segmentation results that aid the following linear model in contrast to predicting from ground truth segmentations, which have been generated by different observers.  
An interesting observation is that a reasonable DICE score is sufficient to achieve excellent performance in diagnostic follow-up tasks. 

A limitation of our study is that we require input images that resemble a 4CH acquisition orientation in a healthy subject. This can be challenging for severely affected patients. However, for cases with severely abnormal hearts, manual detection of CHD would likely be trivial at the point of care, also without segmentation analysis. Good views for borderline cases, which are in focus here, can be identified either manually or with automated view classification~\cite{baumgartner2017sononet}. 

For this work we rigidly aligned all the data to a canonical orientation relative to the heart. This can be achieved in the clinical practice through automated localisation/segmentation/spatial transformer approaches. We observed that this data curation step has a significant impact on all models' performances compared to unaligned images, in which fetuses may present in arbitrary orientation. Accounting for flipped probe orientations paired with hyper-parameter tuning for $\omega, \lambda, \gamma$ would likely lead to further improvements. 

Another limitation is that we do not consider inherent spatio-temporal information of ultrasound imaging. Experienced fetal cardiologists can derive valuable secondary information from how the heart moves. This knowledge can inform future work on the topic. In the clinical practice, still images, as used in our work are common practice to report and document cases, thus a direct application to retrospective quality control and diagnosis support for primary care is in reach.

\section{Conclusion}
We have discussed how segmentation models can be used as clinically interpretable alternative to direct image classification methods for the diagnosis of hypo-plastic left heart syndrome during routine ultrasound examinations. We test a new approach that facilitates disease-status information to bias an automatically constructed atlas label map for robust segmentation and apply Atlas-ISTNs to the problem of fetal cardiac segmentation from ultrasound images for the first time. Our analysis shows that our interpretable approach is en-par with direct image classification, for which ROC-AUC of up to 0.93 is reported~\cite{Tan2020AutomatedScreening}. 
Future work will investigate the true effectiveness of such methods in a prospective clinical trial, which is currently implemented in our clinic.   

\noindent\textbf{Acknowledgements:} We thank the volunteers and sonographers at St. Thomas' Hospital London.
The work of E.C.R. was supported by the Academy of Medical Sciences/the British Heart Foundation/the Government Department of Business, Energy and Industrial Strategy/the Wellcome Trust Springboard Award [SBF003/1116].
We also gratefully acknowledge financial support from the Wellcome Trust IEH 102431, EPSRC (EP/S022104/1, EP/S013687/1), EPSRC Centre for Medical Engineering [WT 203148/Z/16/Z], the National Institute for Health Research (NIHR) Biomedical Research Centre (BRC) based at Guy's and St Thomas' NHS Foundation Trust and King's College London and supported by the NIHR Clinical Research Facility (CRF) at Guy’s and St Thomas’, and Nvidia GPU donations. 

%

\bibliographystyle{splncs04}
\bibliography{ref}
\end{document}